\documentclass[twocolumn,showpacs,preprintnumbers,amsmath,amssymb,pre]{revtex4}

\usepackage{psfig}
\usepackage{bm}
\usepackage{graphicx}

\begin{document}

\setlength{\voffset}{3\baselineskip}

%	\preprint{}

\title{Complex dynamics in simple systems with seasonal parameter oscillations}

\author{L. H\'ector Ju\'arez$^1$, Holger Kantz$^2$, Oscar Mart\'\i nez$^3$,  
 Eduardo Ramos$^3$ and Ra\'ul Rechtman$^3$\\
{\small\it
$^1$Departamento de Matem\'aticas, 
Universidad Aut\'onoma Metropolitana--Iztapalapa,\\
Apdo. Postal 55-534, 09340 M\'exico D.F., Mexico,\\
$^2$Max-Planck-Institut f\"ur Physik komplexer Systeme,
N\"othnitzer Str.\ 38,\\
 D 01187 Dresden, Germany\\
$^3$Centro de Investigaci\'on en Energ\'\i a, Universidad Nacional
Aut\'onoma\\
de M\'exico, Apdo. Postal 34, 62580 Temixco, Mor., Mexico}}

\date{\today}

\begin{abstract}
We study systems with periodically oscillating  
parameters that can give way to complex periodic or non periodic orbits.
Performing the long time limit, we can define ergodic
averages such as Lyapunov exponents, where a negative maximal Lyapunov
exponent corresponds to a stable periodic orbit.  By this, extremely
complicated periodic orbits composed of 
contracting and expanding phases appear in a natural way.  
Employing the technique of $\epsilon$-uncertain points, we find that
values of the control parameters supporting such periodic motion are
densely embedded in a set of values for which the motion is chaotic.
When a tiny amount of noise is
coupled to the system, dynamics with positive and with negative
non-trivial Lyapunov exponents are indistinguishable. We discuss two
physical systems, an oscillatory flow inside a duct and a dripping
faucet with variable water supply, where such a
mechanism seems to be responsible for a complicated alternation of laminar and
turbulent phases.
\end{abstract}


\maketitle

%%%%%%%%%%%%%%%%%%%%%%%%%%%%%%%%%%%%%%%%%%%%%%%%%%%%%%%%%%%%%%%%%%%%%%%%%%%
\section{Introduction}

When studying deterministic dynamical systems, it has become practice to
distinguish between chaotic and (quasi-)periodic solutions, where chaos
has been seen as a novel and strange behavior. We study a class of dynamical
systems, where such a distinction is not useful. Depending on
control parameters, our systems will either generate stable periodic orbits or
they will behave chaotically, but these two types of motion will be essentially
indistinguishable from each other both on computers (due to finite precision)
and in real experiments (due to weak noises).

Seasonal variations are a prominent source of slow periodic
parameter fluctuations in biological, ecological, geochemical and
geophysical systems. But also in many
technical and physical situations, slow periodic oscillations of
system parameters do occur. Speaking of time dependent parameters, we
imply that there is no feedback from the system under study to the
variation of the parameters, whose time dependence can either be
considered as given (non-autonomous situation) or can be ruled by
its own periodic autonomous dynamics. Moreover, we focus on situations
where the typical time scales related to the system dynamics for fixed
parameters is much faster than the time scale related to the parameter
variation, as it is typical of many processes subject to seasonal
variations. The opposite case, where both time scales are comparable has 
been studied as an open loop control mechanism~\cite{mir99,lim90}.

In what follows we discuss a scenario, 
where slow harmonic parameter 
variations introduce an alternation
of expanding and contracting phases. We will show that
ergodic averages can be performed as usual and hence the motion in the
long time limit is clearly classified to be either periodic or
chaotic. However, as our analysis will show, in practical
applications, chaotic motion will be indistinguishable from
periodic motion, both in numerical simulations and in
experiments. In the latter case, this is another example of ``stable
chaos''~\cite{Politi93EPL}. This indistinguishablility implies a
robustness of the phenomenon despite the fact that there are 
parameter regimes where stable periodic motion and chaos are
both supported by a dense set of parameters. 

This rather unexpected behavior is shown to
exist in numerical experiments of an oscillating flow inside a duct and 
a dripping faucet with variable liquid supply.

%%%%%%%%%%%%%%%%%%%%%%%%%%%%%%%%%%%%%%%%%%%%%%%%%%%%%%%%%%%%%%%%%%%%%%%%%%

\section{Map with periodic parameter variations}

Many physical experiments and the corresponding model systems possess
solutions which, due to dissipation, relax to rather uninteresting
fixed points. Such systems are often exposed to a periodic driving, such as
electric resonance circuits. In particular when the system
without driving has a two-dimensional phase space, chaos can only
appear with the driving term.  

This paper deals with a very different class of driven systems; we
assume that our system without driving can behave chaotically,
depending on the values of a control parameter. This parameter is then
varied periodically, with a period which is much longer than the time
scale of the internal dynamics on which the auto-correlation
function in the chaotic regime would relax, or which would govern the
relaxation to a stable fixed point in parameter regimes where it
exists. Hence, the temporal dependence of the control parameter has no
influence on the short time dynamics of the system, but it causes
transitions between different dynamical behaviors 
which might exist for different values of the control parameter.

In what follows we choose the quartic map $f$ given by
\begin{equation}\label{eq:map}
x_{n+1} = f(a,x_n) = 4ax_n(1-x_n)[1-ax_n(1-x_n)]
\end{equation}
with $0<a\leq 4$. This map has the interesting property that for $a=a_c=3.375$
a tangent bifurcation occurs where out of a chaotic region that covers the
whole $[0,1]$ interval a period one window appears. However, other maps can be also 
be considered, for example the logistic map for values of the control parameter
near a period 3 window.

Imposing a temporal dependence on the control parameter leads us to consider the
two dimensional skew autonomous system 
\begin{align}\label{eq:logsyst}
 \phi_{n+1} &= \phi_n + \omega \nonumber\\
 x_{n+1} &= f(a(\phi_n),x_n)\\ 
         &= 4a(\phi_n)x_n(1-x_n)[1-a(\phi_n)x_n(1-x_n)]\nonumber
\end{align}
where $a(\phi_n)=a_0+a_1\cos\phi_n$, $a_0$ is chosen near $a_c$,  $a_1\ll 1$ 
and $\omega=2\pi/N$, $N\gg 1$. A small value of $\omega$
leads to a well defined time scale separation between
the parameter variation and the dynamics of the system while the choice
of $a_0$ and $a_1$ ensures that $a$ oscillates from values where the unperturbed 
map of Eq.~(\ref{eq:map}) is chaotic to values where it has a stable fixed point.

\begin{figure}
 \psfig{file=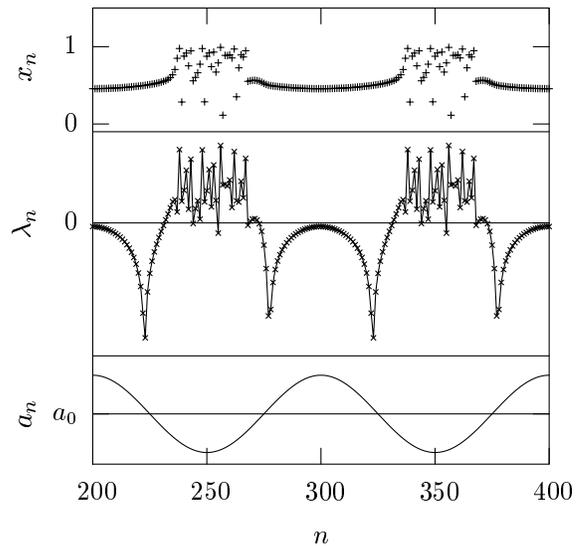}
 \caption[]{\label{fig:map_sequence}\small Two hundred iterates of the map
  Eq.~(\ref{eq:logsyst}) for $a_0=3.4$, $a_1=0.1$, $\omega=2\pi/100$ (top panel),
  the instantaneous stretching factors $\lambda_n$ (middle panel)
  and the parameter $a_n$ (bottom panel).}
\end{figure}

A typical sequence of iterates of Eq.~(\ref{eq:logsyst}) (after discarding
a transient) is shown in the top panel of Fig.~\ref{fig:map_sequence}. We see an
alternation between irregular fluctuations of $x$ and regular
episodes due to the oscillating values of $a$ shown in the bottom panel. 
Due to the slowness of the change of $a$, the trajectory
can relax toward the stable fixed point for those values of $a_n$
where it exists, whereas it follows an irregular trajectory for
$a_n \lesssim 3.375$, where the map of Eq.~(\ref{eq:map}) is chaotic.

Since the $\phi$-dynamics is not mixing, this system
clearly has (at least) one invariant measure in its two-dimensional
phase space for every $\phi_0$, on which ergodic averages are well
defined. The Jacobi matrix of Eq.(\ref{eq:logsyst}) is triangular,
which immediately shows 
that one Lyapunov exponent (corresponding to the $\phi$-dynamics) is zero,
whereas the other $\lambda=\langle\lambda_n\rangle$ is found 
as an average over an infinitely long trajectory of the instantaneous stretching 
factors $\lambda_n$ (middle panel of Fig.~\ref{fig:map_sequence} defined by
\[
 \lambda_n = \log\left|\dfrac{\partial f(a(\phi_n),x_n)}{\partial x_n}\right|.
\]
Assuming ergodicity, $\lambda$ can also be found as an average of the
stretching factor over the invariant measure.
The seemingly intermittent dynamics shown in in the upper panel of 
Fig.~\ref{fig:map_sequence} has a well defined non-trivial Lyapunov exponent, 
whose sign depends on the details of the $x$-dynamics. The alternation between
unstable and stable phases is different from the typical intermittency
scenarios~\cite{Maneville}. The contributions of the stable and
unstable phases  
to the Lyapunov exponent have a sensitive and subtle dependence on the choice 
of $a_0$ and $a_1$ as we show in Sec.~\ref{par-dependence}.

%%%%%%%%%%%%%%%%%%%%%%%%%%%%%%%%%%%%%%%%%%%%%%%%%%%%%%%%%%%%%%%%%%%%%%%%%%%%%

\section{Indistinguishability between chaotic and non chaotic orbits}

For the map of Eq.~(\ref{eq:map}) almost any initial condition with $a>a_c$,
$|a-a_c|\ll 1$ settles after a transient on a fixed point $\hat{x}(a)$.
On the other hand, for  $a<a_c$, $|a-a_c|\ll 1$, the invariant set is 
one-dimensional for a dense set of parameter values, but shows type-I 
intermittency because of the closeness to the tangent bifurcations. 
For the map of Eq.~(\ref{eq:logsyst}) during one period of the auxiliary 
variable $\phi_n$, the parameter $a(\phi_n)$ alternates between the periodic and 
chaotic regime of the map of Eq.~(\ref{eq:map}) and therefore $\lambda$ has 
contributions with negative and positive signs. Depending on which contribution 
has a larger modulus, the overall dynamics is either chaotic or not. 
This in turn, depends on the values of $a_0$, $a_1$ and $\omega$.

During the iterates where the trajectory looks regular and is near to
$\hat{x}(a)$, the fixed point of the map of Eq.~(\ref{eq:map}), 
the tangent space dynamics is contracting, and we call the
accumulated contraction factor $f_c$. During the iterations when the
trajectory looks irregular, its tangent space dynamics is essentially
expanding, and we call the accumulated expansion factor $f_e$. Then,
roughly, the Lyapunov exponent is $\lambda \approx \langle \ln |f_c|
+ \ln |f_e|\rangle$, where $\langle\cdots\rangle$ now denotes the
average over successive periods of $\phi$. If we start two
trajectories with a distance $\epsilon$ at the beginning of the
irregular phase, at its end their distance is $\epsilon f_e$. This
distance will shrink during the regular phase, and
at its end will be $\epsilon f_e f_c$. Hence, if $f_e f_c$ on
average is smaller than unity, two trajectories will approach each
other, and finally will be indistinguishable.

This alternation between contracting and expanding episodes
is clearly visible in Fig.~\ref{fig:map_sequence}. A stable periodic
orbit thus is as irregular as a chaotic solution, but its irregular
segment repeats itself exactly in every period of oscillation of the
parameter $a$, whereas for a trajectory with a positive Lyapunov
exponent it does not. Hence, this system
can create arbitrarily complicated periodic orbits, since by the
choice of $\omega$ one can determine the period length and also, how
many points of the periodic orbit are in the irregular regime. When
the $\phi_n$ dynamics is quasi-periodic instead of periodic (by
choosing $N$ as an irrational number), also
orbits with negative Lyapunov exponent have non-repeating irregular
segments. By visual inspection, these cannot be distinguished from
orbits with a positive Lyapunov exponent.

%%%%%%%%%%%%%%%%%%%%%%%%%%%%%%%%%%%%%%%%%%%%%%%%%%%%%%%%%%%%%%%%%%%%%%%%%%%%%%%

\section{The stroboscopic view}

\begin{figure}
 \psfig{file=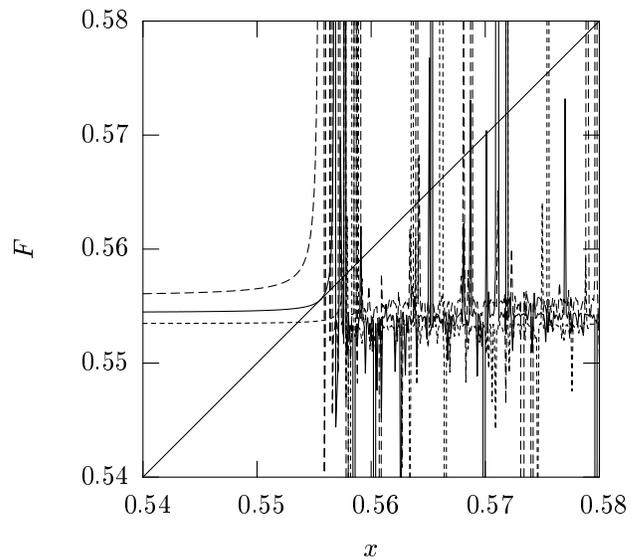}
 \caption[]{\label{fig:graph}\small The graph $F$ as defined
  in the text for $a_0=a_c=3.375$ (full curve) showing a
  tangent bifurction, for $a_0=a_c-0.0005$ (dashed curve) showing many
  unstable fixed points and for $a_0=a_c+0.0005$ (dotted curve) showing an
  almost super stable fixed point.}
\end{figure}

The special dynamics of $\phi_n$ hinders the 2-dimensional map of
Eq.~(\ref{eq:logsyst}) of having a fixed point. The shortest periodic
orbit can have length $N$ when $\omega=2\pi/N$.  Therefore, it makes sense
to study the composition of $N$ successive iterates 
$F(x)=\bigcirc_{n=1}^N f(a(\phi_n),x_n)$. When the Lyapunov exponent of
Eq.~(\ref{eq:logsyst}) is negative, $F$ should have a stable
periodic orbit or fixed point, whereas it has only unstable periodic
orbits and chaotic solutions for positive exponents. 
In Fig.~\ref{fig:graph} we show the graph of $F$ for several values of $a_0$.
For $a>a_c$, $a-a_c\ll 1$ $F(x)$
 has an almost super-stable fixed
point. These fixed points are generically born and eliminated by
tangent bifurcations, together with their unstable counterparts.
For every initial phase $\phi_0$ we have a different stroboscopic map $F(x)$.
However they are all topologically conjugate. For our
choice of the variation of $a(\phi_n)$, however, the existence or
absence of a stable fixed point can best be seen when $\phi_0\approx 0$,
where each orbit of the system Eq.~(\ref{eq:logsyst}) assumes values
close to the fixed point of $f(x,a)$ in Eq.~(\ref{eq:map}) 
for fixed $a\approx a_0+a_1$.

%%%%%%%%%%%%%%%%%%%%%%%%%%%%%%%%%%%%%%%%%%%%%%%%%%%%%%%%%%%%%%%%%%%%%%%%%%%%%%%%%%

\section{Noise and round-off effects}

The results described above, and in particular the distinction between
motion corresponding to negative and positive Lyapunov exponents, is
correct only in the abstract mathematical setting. On a computer, the
finite precision of the internal representation of real numbers can
enforce the motion onto a complex periodic orbit although its Lyapunov
exponent is positive. This happens, when 
the contraction factor $f_c$ during a contracting phase is too strong, 
so that at the end of this phase the trajectory has no memory of the
previous expanding phase.
For typical values of $a_0$ and $a_1$ we found numerically that orbits with 
a positive Lyapunov exponent became periodic when the contracting 
phase contained more than thirty 
iterates. If these phases are shorter, either because the period $N$ 
is small enough, or because $a_0$ and $a_1$ are chosen to be inside the 
chaotic regime, orbits with positive Lyapunov exponents are non--periodic 
as expected. Hence, with
finite precision and large $N$, one cannot decide, without computation
of the Lyapunov exponent, whether the system has a stable periodic
orbit or not.

In an experimental realization, instead of computer
round-off errors there is external noise coupled into the
system. Eq.~(\ref{eq:map}) has then to be modified by adding white noise
$\sigma\xi_n$, where $0<\sigma\ll 1$, 
$\langle \xi_n \xi_{n'}\rangle = \delta_{n,n'}$ and 
$\langle\xi_n\rangle = 0$. This has no
visible effect on chaotic solutions of Eq.~(\ref{eq:logsyst}), but
it does destroy the periodicity of stable periodic solutions. Inside
the expanding and hence irregular sections, noise is exponentially
amplified and creates orbits which appear chaotic. Systems like
Eq.~(\ref{eq:logsyst}) therefore have periodic orbits which are
extremely sensitive to external noise, despite linear stability. 

%%%%%%%%%%%%%%%%%%%%%%%%%%%%%%%%%%%%%%%%%%%%%%%%%%%%%%%%%%%%%%%%%%%%%%%%%%%%%%%%%%%

\section{Parameter dependence}
\label{par-dependence}

\begin{figure}
\psfig{file=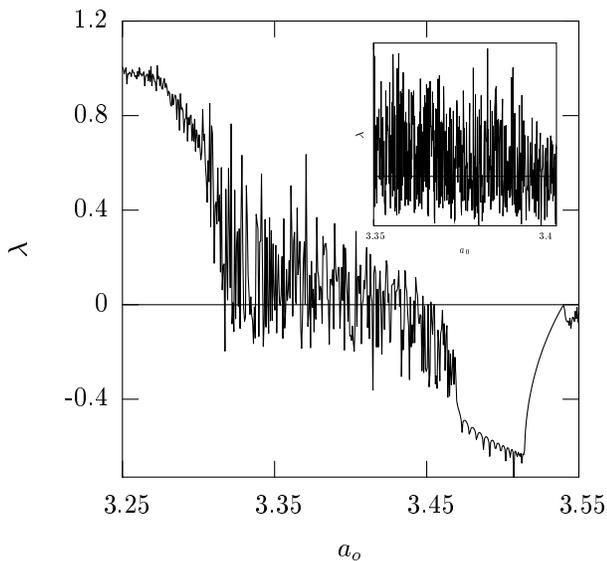}
\caption[]{\label{fig:lyaps}\small The Lyapunov exponent $\lambda$ as a function 
 of the parameter $a_0$  with $a_1=0.1$, $N=100$. The inset shows a smaller 
 interval of $a_0$ at a higher resolution.} 
\end{figure}

\begin{figure}
\psfig{file=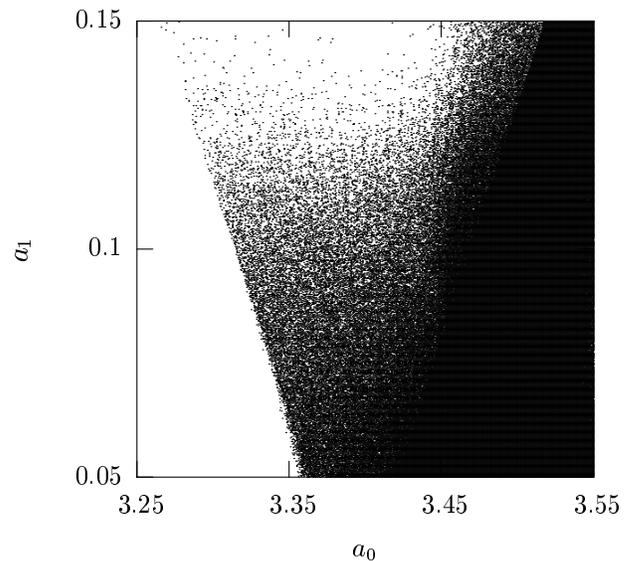}
 \caption{\label{fig:pf-9-y-100} Lyapunov diagram with $N=100$. Black dots 
  correspond to a negative Lyapunov exponent, white to a positive one. A total of 
  65,536 dots are drawn. A cut at $a_1=0.1$ corresponds to the previous Fig.}
\end{figure}

\begin{figure}
\psfig{file=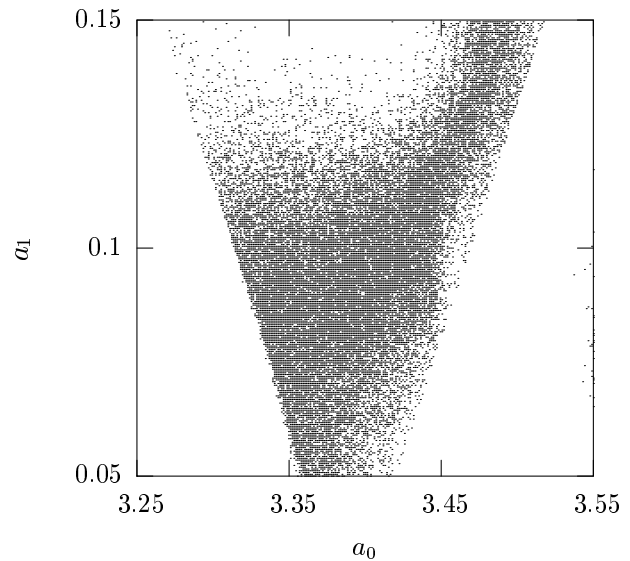}
 \caption{\label{fig:pf-8-y-100} Black dots represent the $\epsilon$--uncertain 
  points with $\epsilon=10^{-7}$, and $N=100$.}
\end{figure}

The rather complex sequence of tangent bifurcations leading to the
creation and destruction of the stable periodic orbits 
causes the orbits to depend sensitively on the system parameters, so
that there is a complicated flipping from periodic to chaotic motion
as a function of every single parameter as illustrated in 
Figs.~\ref{fig:lyaps}--\ref{fig:pf-8-y-100}. In the first one, we show
the Lyapunov exponent $\lambda$ as a function of $a_0$ 
($a_1$ and $N$ fixed). The rapid change from chaotic to stable periodic 
solutions leads us to 
assume that both types of behaviors are supported by a dense set of parameters.
The same follows from Fig.~\ref{fig:pf-9-y-100} where we show the 
Lyapunov diagram where white (black) dots correspond to a
positive (negative) value of $\lambda$~\cite{mar89}. There are clearly two regions
where $\lambda$ has a definite sign and an intermediate one where slight
changes of $a_0$ or $a_1$ have a dramatic effect on the dynamics.
We determined 
the fractal dimension of the boundary between stable and unstable 
solutions by finding the scaling of the number of $\epsilon$--uncertain 
points as $\epsilon$ is varied~\cite{ble88,ott93}. A point $(a_0,a_1)$ 
is $\epsilon$--uncertain if in a neighorhood of radius $\epsilon$ 
there exists at least one point where the Lyapunov exponent has an 
opposite sign to the one evaluated at  $(a_0,a_1)$. In 
Fig.~\ref{fig:pf-8-y-100} we show those points of the previous figure that 
are $\epsilon$--uncertain for $\epsilon=10^{-7}$. From the scaling of the 
number of uncertain points with $\epsilon$ we found the box counting 
dimension $D_0$ of the boundary between chaotic and stable solutions to 
be $D_0\sim 1.985$. This result gives a quantitative support to the idea 
that both stable and unstable orbits have a dense set of parameters. We also 
found that $D_0\sim 2$ for $N>100$ and also for $N<20$ while it has a minumum 
value $D_0\sim 1.8$ for $N\sim 30$. 

%%%%%%%%%%%%%%%%%%%%%%%%%%%%%%%%%%%%%%%%%%%%%%%%%%%%%%%%%%%%%%%%%%%%%%%%%%%%%%%%%%

\section {Oscillatory flow in a duct}\label{sec:duct}

As a first example that displays the behavior discussed
previously we briefly present results of numerical experiments of
an oscillatory flow in a duct filled with fluid. This flow can be generated by 
imposing oscillatory pressure or velocity fields at the ends of the duct with 
suitably defined phase lags. The stability of these flows can be described in 
terms of two nondimensional parameters, the oscillatory Reynolds number 
$R_{\delta}$ and the Stokes parameter $\lambda$. These are defined by 
$R_{\delta}= U\delta/\nu$ and $\lambda=D/\delta$ where $U$ is a characteristic 
velocity, $\nu$ the kinematic viscosity of the fluid, $D$ a characterisic distance 
of the duct and $\delta$ the Stokes penetration depth. This last quantity is 
defined by  $\delta=\sqrt{\nu/2\omega}$ with $\omega$ the frequency of the
oscillation. 
It is a well established fact that zones of distinct dynamical qualitative behavior
can be identified in the $(R_{\delta}$,$\lambda)$ space. Specifically, it has
been experimentally observed that for
$\lambda \ge 2$ and $R_{\delta} < 500$, the flow is laminar, while for
$R_{\delta} > 500$, the flow inside the duct is laminar for the phase
intervals where the velocity is small while bursts with a frequency much
larger than the forcing appear near the end of the acceleration phase~\cite{Hino}.
As $R_{\delta}$ is increased, the phase interval where high frequency
oscillations are present gets larger, but it never covers the cycle
entirely. 
The origin of the high frequency oscillations is the generation of
vortices due to the instability of the laminar flow. Numerical results 
agree with the experiment~\cite{Juarez}. 

\begin{figure}
 \psfig{file=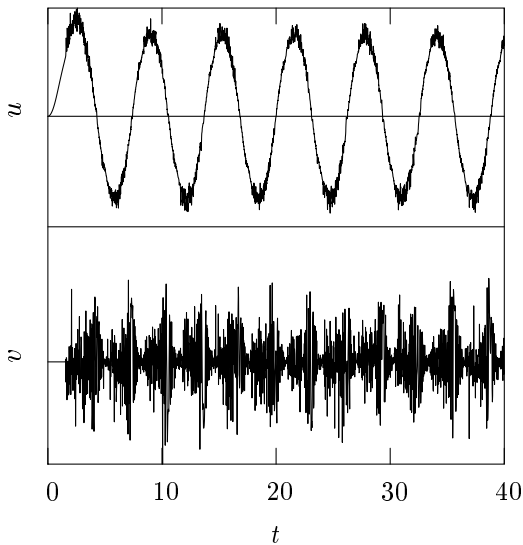}
 \caption[]{\label{juarez1500}\small Velocities $u$ and  $v$ as a function of 
  time $t$ in arbitrary units at the center of a two dimensional duct with 
  expansions at the ends and with an aspect ratio (total length/cross section) 
  of 20, $R_{\delta}= 1521$ and $\lambda \sim 4$.}
\end{figure}

In particular, Fig.~\ref{juarez1500}
shows the axial $u$ and transversal $v$ velocities in the middle of a duct.
As can be observed, when the
velocity is close to zero, the trace is smooth, indicating laminar
flow, however, high frequency oscillations appear when the velocity reaches its
maximum absolute value in each cycle.
Taking $\omega_v = U/D$, as the lower limit of the characteristic 
frequency of the vortices, the ratio of the vortices frequency to
the forcing frequency, $\omega_v /\omega $ is $2 R_{\delta}/\lambda$ which
for this example is 768. This indicates that the changes in the driving 
force are slow compared with the internal vortex dynamics corresponding
with the conditions discussed in the previous sections. 

%%%%%%%%%%%%%%%%%%%%%%%%%%%%%%%%%%%%%%%%%%%%%%%%%%%%%%%%%%%%%%%%%%%%%%%%%%%%%%%%%%

\section{Dripping faucet with variable supply}

Our second example describes the dynamics of a dripping faucet with
variable liquid supply. This problem, with constant liquid supply has been 
studied extensively~\cite{Shaw}. The time interval
$T$ between successive drops shows a complicated bifurcation diagram 
as the water supply $\epsilon$ increases.  There are two
studies that are important in the context of the present analysis. The model
presented by Fuchikami \textit {et al}~\cite{Fuchikami} is relevant because 
it is built on sound physical phenomenology, but unfortunately is not simple and 
long time calculations involving many drops are extremely computing
intensive. On the other hand, the model presented by Coullet 
\textit {et al}~\cite{Coullet} which is based on the former, is 
useful since it is simple and can be used for exploring the long time behavior. 
Both models can be adapted to
analyze the system when the liquid supply varies as a harmonic function
of time and they display a similar behavior with 
constant and variable liquid supply. 

\begin{figure}
 \psfig{file=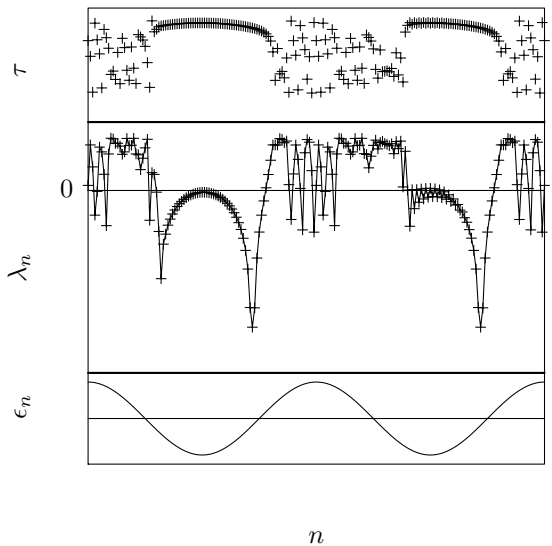}
 \caption{\label{fig:coullet2} Two hundred iterates of the map of Coullet
  {\it et al} mentioned in the text (top panel), the instantaneous
  stretching rates $\lambda_n$ (middle panel) and $\epsilon$ varying 
  harmonically (bottom panel)  around $\epsilon_0=0.0146969$ with an 
  amplitude $\epsilon_1=0.0004$ and a frequency $\omega=2\pi/100$.}
\end{figure}

The Coullet model with constant liquid supply $\epsilon$ has a complicated 
bifurcation diagram and we choose a value $\epsilon_c$ in such a way that
to its left there is are period one solutions, and to its right a densely
chaotic region.
We now consider a variable liquid supply that varies harmonically
around a value near $\epsilon_c$
with a period which is approximately 1000 times larger than 
the characteristic period of the subsequent drop release with 
constant liquid supply and with an amplitude $\Delta\epsilon$
chosen in a way that the water supply does not extend outside the 
period four window.
These conditions allow the system to visit alternatively a zone of irregular 
behavior and a zone of period four in the map of constant liquid supply.
The result of the periodic variation of the water supply are shown in 
Fig.~\ref{fig:coullet2}.
The continuous sinusoidal line represents the liquid supply and the dots
the values of the time intervals between successive drops.
As can be clearly seen, the system presents zones of stable and unsatble behavior
as for the map of Eq.~(\ref{eq:logsyst}).    

%%%%%%%%%%%%%%%%%%%%%%%%%%%%%%%%%%%%%%%%%%%%%%%%%%%%%%%%%%%%%%%%%%%%%%%%%%%%%%%%%%%

\section{Conclusions}
We discussed systems with periodic parameter fluctuations which are
driven from regular to chaotic motion and back. Although
simple in its construction, this type of dynamics creates very
complicated orbits with a complex
dependence on control parameters. 
In computer simulations and in real experiments, it
is impossible to distinguish the existence of stable complex periodic
solutions and of chaotic solutions in the underlying model, since
round-off errors and noise interact with the dynamics.

We presented two numerical experiments that illustrate the change of behavior 
due to a periodic variation of a parameter with a time scale much larger than 
the natural time scale of the system where the discussion of the impossibility
of distinguishing chaos from order is relevant.

Our detailed analysis was based on maps, but all features are found as
well in systems with continuous time.
It is also not relevant to assume sinusoidal variation of the
parameter, so that we expect such a behavior to be rather widespread.

%%%%%%%%%%%%%%%%%%%%%%%%%%%%%%%%%%%%%%%%%%%%%%%%%%%%%%%%%%%%%%%%%%%%%%%%%%%%%%%%%%%

\section*{Acknowledgments}

Assistance of H\'ector Cort\'es and Alfredo Quiroz with numerical calculations is 
gratefully acknowledged. 
This work has been partially supported by DGAPA--UNAM under Grant No.~IN103300. 

%%%%%%%%%%%%%%%%%%%%%%%%%%%%%%%%%%%%%%%%%%%%%%%%%%%%%%%%%%%%%%%%%%%%%%%%%%%%%%%%%%%

{}

\end{document}